\documentclass[a4paper, 11pt]{article}
\pdfoutput=1
\usepackage{jcappub}
\usepackage[utf8]{inputenc}
\usepackage{graphicx}
\usepackage{amsmath}
\usepackage{amsfonts}
\usepackage{amssymb}
\usepackage{bbold}
\usepackage{float}

\usepackage{tikz}
\usetikzlibrary{arrows,backgrounds,snakes,patterns}
\usetikzlibrary{shapes,arrows,chains}
\usepackage{verbatim}
\usepackage{booktabs}
\usepackage{pgfplots}
\pgfplotsset{compat=1.10}
\usepgfplotslibrary{fillbetween}
\usepackage[flushleft]{threeparttable}
\graphicspath{ {./figures/} }
\usepackage{array}
\usepackage{placeins}

\usepackage{subcaption}
\usepackage{datetime2}
\usepackage{physics}

\newcommand{\cG}{{\cal G}}
\newcommand{\cH}{{\cal H}}
\newcommand{\cK}{{\cal K}}
\newcommand{\cF}{{\cal F}}
\newcommand{\cD}{{\cal D}}
\newcommand{\cM}{{\cal M}}
\newcommand{\cN}{{\cal N}}

\newcommand{\cP}{{\cal P}}
\newcommand{\cO}{\mathcal O}

\newcommand{\Bf}{\bar{\phi}}

\newcommand{\Mpl}{\ensuremath M_{\textrm{Pl}{}}}

\numberwithin{equation}{section}
\numberwithin{figure}{section}

\title{Primordial Stochastic Gravitational Wave Backgrounds from a Sharp Feature in Three-field Inflation II: The Inflationary Era}

\author[a]{Vikas Aragam}
\author[b]{Sonia Paban}
\author[c,d]{Robert Rosati}

\affiliation[a]{
        Medical Physics Division, Thomas Jefferson University, Philadelphia, PA 19107, USA
}

\affiliation[b]{
	Jefferson Physical Laboratory, Harvard University, Cambridge, MA 02138, USA

	}

\affiliation[c]{
	NASA Postdoctoral Program Fellow, NASA Marshall Space Flight Center, Huntsville, AL 35812, USA
}
\affiliation[d]{
	Department of Astronomy and Astrophysics, University of California, Santa Cruz, CA 95064, USA
}
\emailAdd{aragam@utexas.edu}
\emailAdd{sppaban@fas.harvard.edu}
\emailAdd{robert.j.rosati@nasa.gov}

\abstract{
We study the contribution of large scalar perturbations sourced by a sharp feature during cosmic inflation to the stochastic gravitational wave background (SGWB), extending our previous work to include the SGWB sourced during the inflationary era.
We focus in particular on three-field inflation, since the third dynamical field is the first not privileged by the perturbations' equations of motion and allows a more direct generalization to $\mathcal{N}$-field inflation.
For the first time, we study the three-field isocurvature perturbations sourced during the feature and include the effects of isocurvature masses. 
In addition to a two-field limit, we find that the third field's dynamics during the feature can source large isocurvature transients which then later decay, leaving an inflationary-era-sourced SGWB as their only observable signature.
We find that the inflationary-era signal shape near the peak is largely independent of the number of dynamical fields and has a greatly enhanced amplitude sourced by the large isocurvature transient, suppressing the radiation-era contribution and opening a new window of detectable parameter space with small adiabatic enhancement. The largest enhancements we study could easily violate backreaction constraints, but much of parameter space remains under perturbative control. These SGWBs could be visible in LISA and other gravitational wave experiments, leaving an almost universal signature of sharp features during multi-field inflation, even when the sourcing isocurvature decays to unobservability shortly afterwards.
}
\arxivnumber{2409.09023}

\begin{document}
\maketitle
\thispagestyle{empty}
\newpage

\setcounter{page}{1}

\pgfmathsetmacro{\T}{1.0} 
\pgfmathsetmacro{\O}{2.0} 
\pgfmathsetmacro{\D}{1.0} 

\section{Introduction}

 Exploring the stochastic gravitational wave background (SGWB) could unlock invaluable information about astrophysical source populations and processes that remain out of reach through any other means. Gravitational waves can, for example, give us information all the way back to the onset of inflation \cite{Maggiore_2000,Romano:2016dpx,Domenech:2021ztg}.  But to learn from this complex data, we must have precise predictions for the observables commonly used to characterize it: fractional energy density spectrum ($\Omega_{\rm GW}$), characteristic strain, and distribution of gravitational wave power on the sky. 

  Inflation is among the cosmological sources of SGWB. Single-field slow-roll inflation, without features, predicts an $\Omega_{\rm GW}$ that is almost constant in frequency and whose magnitude is well below the sensitivity level of future detectors \cite{Turner:1996ck}. Single-field models with periods of ultra-slow rolling or rapid-turning multi-field models of inflation, however, can source detectable gravitational waves at second order in perturbation theory \cite{Ananda:2006af, Zhou:2020kkf, Fumagalli:2020nvq, Braglia:2020taf, Fumagalli:2021mpc, Bhattacharya:2022fze, Braglia:2022ftm, Dimastrogiovanni:2023oid,Braglia:2024kpo}.  In  \cite{Fumagalli:2021mpc}, Fumagalli {\it {\it et al.}} derive the tensor power spectrum sourced at second order from excited states during inflation in a generic multi-field model. One concrete realization of this scenario, which can occur in SUGRA models \cite{Aragam:2021scu, Anguelova:2020nzl}, is a brief rapid turn in field space. This turning is a departure from single-field behavior and sources both a feature in ${\cal P}_{\zeta}$ (at shorter scales than observed by CMB and LSS)  and the Stochastic Gravitational-Wave Background (SGWB).  The authors of \cite{Fumagalli:2021mpc} compute analytically and numerically the frequency profile of $\Omega_{\rm GW}$ for two fields in this scenario.  Then, they hypothesize that if the contributions from all the fields are of the same order, the power spectrum is enhanced by a ${\cal N}^4$ factor, where ${\cal N}$ is the number of fields. 

In this work, we examine and extend the findings of Fumagalli {\it et al.} \cite{Fumagalli:2021mpc} to identify the conditions under which all fields contribute equally. Specifically, we calculate the contributions to $\Omega_{\rm GW}$ in three-field models by varying the relative values of the turn rate and torsion. This allows for a smooth transition from two to three active dynamic fields. 
In the evolution of the fluctuations, ${\cal N}=3$ is distinct because for ${\cal N} \geq 3$, only one linear combination of isocurvature fluctuations couples directly to the adiabatic fluctuations, enabling a more direct generalization to ${\cal N}$  fields. Enhanced power spectra and primordial black holes from broader features with three fields were also studied in \cite{Christodoulidis:2023eiw}.

Secondary gravitational waves are generated at two times: during inflation, $\Omega^{\rm{inf}}_{\rm GW}$ and during the radiation-dominated era as the features in the scalar spectrum reenter the horizon $\Omega^{\rm{rad}}_{\rm GW}$. In \cite{Aragam:2023adu}, we examined the ${\cal N}$ dependence in the latter case, finding the frequency profile to depend on the number of active fields, and hinted that not all fields may contribute equally to $\Omega_{\rm GW}.$ 

When the mechanism that generates excited states is a brief turn in field space, the mode that crosses the horizon when the turn occurs becomes a reference scale $k_f$.
Fumagalli {\it et al.} \cite{Fumagalli:2021mpc} demonstrated that, for ${\cal N}=2$, the frequency profile of $\Omega^{\rm{inf}}_{\rm GW}$ has a principal peak at  $k_{\rm{max}} \simeq {\cal O}(1) k_{f}, $ followed by a series of order-one oscillations with frequency  $2/{k_f}$. Our findings indicate that the location of the maximum ($\sim k_f$) and the frequency of the oscillation  ($2/{k_f}$), as derived in  \cite{Fumagalli:2021mpc}, are the same in the two- and three-field limits. By contrast, the $\Omega^{\rm{rad}}_{\rm GW}$ envelope depends on ${\cal N}$ \cite{Aragam:2023adu}. The $\Omega^{\rm{inf}}_{\rm GW}$ peak enhancement does not scale as ${\cal N}^4$ , but it is notably larger than for ${\cal N}=2$, and is primarily caused by a large isocurvature transient.  Our analysis reveals that isocurvature modes for low $k$ grow rapidly when torsion is comparable to, or greater than, the turning. Therefore, in order to match the unobserved isocurvature in CMB data, we assign them a mass of ${\cal O}(H)$. This has been an implicit assumption in previous literature on sharp features. We have, for the first time, calculated all the massive Bogoliubov coefficients analytically and demonstrated that although the isocurvature perturbations decay quickly while the modes are beyond the horizon, they still leave an imprint on the SGWB that could be observable in future gravitational wave experiments.

The presentation of our work is organized as follows. In Section \ref{sec:The set-up}, we review the three-field inflation formalism, summarizing the concepts of torsion and turning rate of background motion and their effect on the time evolution of the linear quantum perturbations. In Subsection \ref{subsect: The tensor power spectrum}, we recap how to compute the contribution of the scalar perturbations to the tensor power spectrum. The results are presented in Section \ref{sec: Results}. The unexpected considerable growth in the perturbations demands a critical appraisal of the results we present in Section \ref{sec:Backreaction}. In Section \ref{sec:Phenomenology}  we discuss the phenomenology of sharp features in CMB, LSS and SGWB.

\section{The Set-up}
\label{sec:The set-up}

In this section, we give a self-contained summary of the set-up. We refer the reader to \cite{Fumagalli:2021mpc} for the general derivation of the contribution to $\Omega_{\rm GW}$ from excited states and to \cite{Christodoulidis:2022vww,Aragam:2023adu,Christodoulidis:2023eiw} for more details on the three field dynamics. 

\begin{figure}
    \centering
    \includegraphics[width=\textwidth]{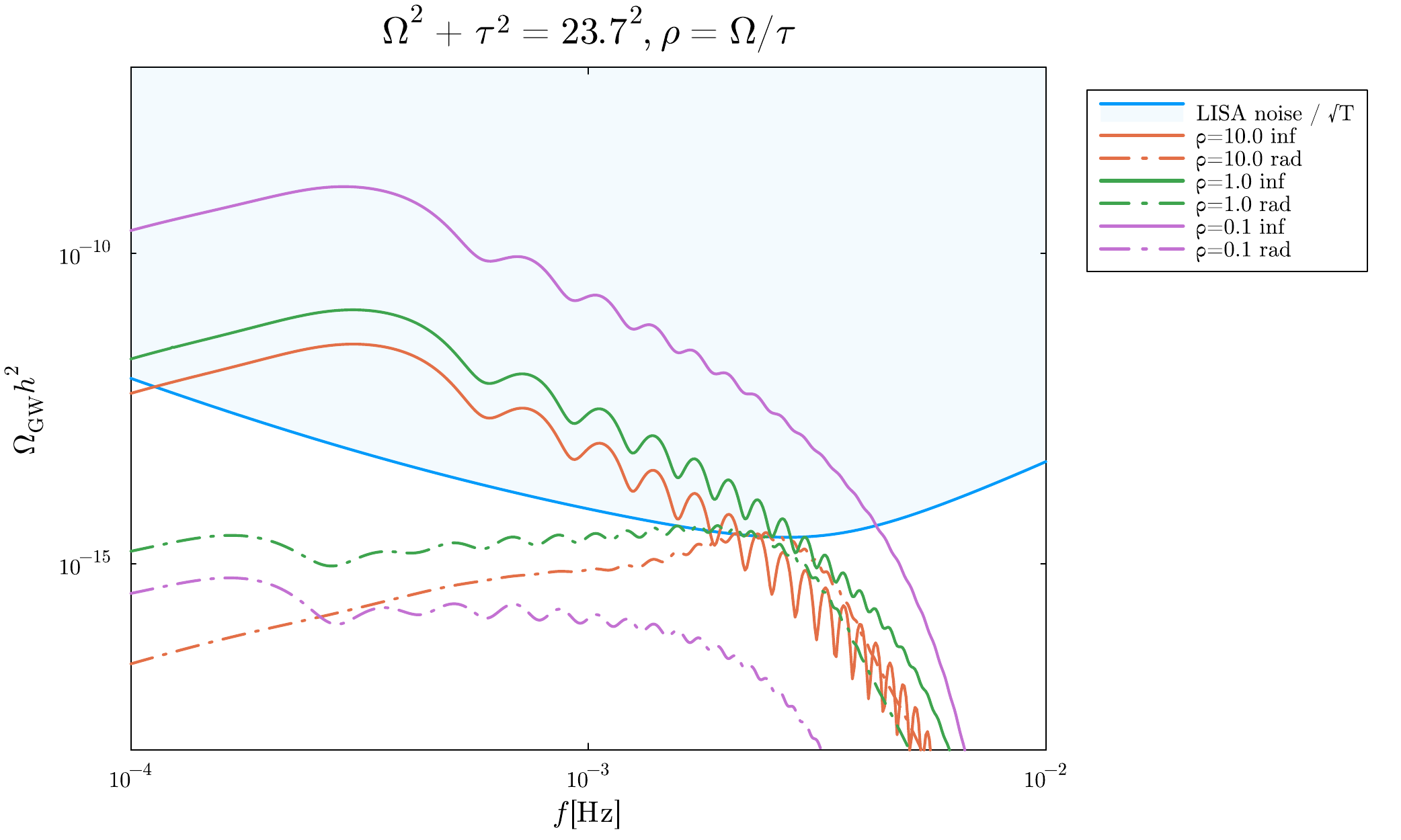}
    \caption{The stochastic gravitational-wave backgrounds computed in this work. We show the range of possibilities for the spectrum in three fields, fixing the approximate effective two-field enhancement of the curvature perturbation ($\Omega_\mathrm{2f}^2 \equiv \Omega^2 + \tau^2$) but varying the ratio of turning and torsion $\rho\equiv\Omega/\tau$. The inflationary-era background is shown in solid lines, while the corresponding radiation-era background is shown in the same color as dot-dashed, with the predicted observable signal being the sum of the two. As we discuss below, when $\rho \lesssim 1$, the isocurvature power spectra are large at horizon-exit (and then rapidly decay) and this can enhance the inflationary-era SGWB without substantially enhancing the curvature perturbation $P_\zeta$ or the radiation-era signal. We therefore expect highly dimensional field spaces to source these inflationary-era SGWBs much more easily than the radiation-era ones.}
    \label{fig:inflationary-sgwb}
\end{figure}

\subsection{Background Motion}

We consider scenarios where inflation is driven by three scalar fields minimally coupled to gravity. The action is:

\begin{equation}
    S= \int \, d^4 x \, \sqrt{-g} \left[ \frac{\Mpl^2}{2} R (g) -\frac{1}{2} \cG_{ab}\, g^{\mu \nu}\, \partial_{\mu} \phi^a\partial_{\nu} \phi^b -V(\phi^a)  \right] \label{eq:action}
\end{equation}  Greek letters label spacetime indices, and lower Latin indices label field-space indices, $a=1,2,3$. $g^{\mu \nu}$ is the spacetime metric and $\cG_{ab}$ is the field space-metric.  $\Mpl=\sqrt{8 \pi G}$ is the reduced Planck mass. We will work in units where  $\Mpl=1.$
 
This work uses the kinematic field basis, so called because it is defined from the fields' trajectory.
The first unit vector in the basis is the velocity unit vector, $\hat{\sigma}^a \equiv \dot{\bar{\phi}}^a/\dot{\bar{\phi}}$, and subsequent unit vectors are defined by additional covariant time derivatives as summarized in the Frenet-Serret system
\begin{align}
\cD_N
\begin{pmatrix}
\hat{\sigma}^a\\
\hat{s}^a\\
\hat{b}^a
\end{pmatrix}
&= \begin{pmatrix}
0 & \Omega & 0 \\
-\Omega & 0 & \tau \\
0 & -\tau & 0
\end{pmatrix}
\begin{pmatrix}
\hat{\sigma}^a\\
\hat{s}^a\\
\hat{b}^a
\end{pmatrix} \equiv \Omega^a_b 
\begin{pmatrix}
\hat{\sigma}^b\\
\hat{s}^b\\
\hat{b}^b
\end{pmatrix},
\label{eq:frenet-serret}
\end{align}
where  $\cD_N A^a \equiv (d{A}^a/dN)+ \Gamma^a_{bc} A^b (d{\Bf}^c/dN)$, and $N$ is the number of e-folds. $\Omega$ and $\tau$ measure the turn rate and torsion of the trajectory respectively, in agreement with the literature \cite{Pinol:2020kvw,Christodoulidis:2022vww}.
When $\Omega >0$, the trajectory undergoes turning, and when $\tau > 0$ as well, that turning is non-planar. If both $\Omega$ and $\tau$ are constant, the trajectory follows a helix. In general,  $\tau$ and the kinematic basis are not well-defined when $\Omega = 0$. A schematic representation of the trajectory during such a feature is available in Figure \ref{fig:turn_profile}.

\subsection{Perturbations}
The linear equations of motion for the Mukhanov-Sasaki variables in kinetic basis $Q_a \equiv \{ Q_\sigma,Q_s,Q_b\}$ can be written as \cite{Aragam:2023adu, Pinol:2020kvw}
\begin{equation}
\begin{aligned}
\cD_N (Q^\prime)^a &+ F^a_b (Q^\prime)^b + C^a_b Q^b = 0 \\ \\
F^a_b &\equiv (3-\epsilon)\delta^a_b - 2\Omega^a_b \\ \\
C^a_b &=  \left(\frac{k}{a H}\right)^2 \delta^a_b  + \begin{pmatrix}
0 & -2(3-\epsilon)\Omega & 0 \\
0 & \cM_{ss} - \Omega^2 -\tau^2  & \cM_{sb} - \tau (3-\epsilon) \\
0 & \cM_{sb} + \tau (3-\epsilon) & \cM_{bb} - \tau^2
\end{pmatrix} + \cO(\epsilon^2,\eta,\nu,\nu_\tau)
\label{eq:kinetic_pert_eom}
\end{aligned}
\end{equation}
The prime denotes a derivative with respect to conformal time.
 $\epsilon \equiv -\cH'/\cH$, $\eta \equiv \epsilon'/\epsilon$, $\nu \equiv \Omega^\prime / \Omega$ and  $\nu_\tau \equiv \tau^\prime / \tau$.  $\{\cM_{ss}, \cM_{sb}, \cM_{bb}\}$ are defined as contractions with $\{\hat{s}^a, \hat{b}^a\}$ of:

\begin{equation}
     \mathcal{M}_{ab} \equiv \frac{V_{;ab}}{H^2}  - 2 \epsilon R_{a\sigma \sigma b} + 2\epsilon(3-\epsilon)\hat{\sigma}_a \hat{\sigma}_b +\sqrt{2\epsilon}\frac{\hat{\sigma}_a V_{,b} + \hat{\sigma}_b V_{,a}}{H^2 \Mpl^2}
    \label{eq:pert_lagrangian}
\end{equation}
To simplify ${C_a}^b$, we have used that some mass matrix elements can be expressed purely in kinematic quantities using the background equations of motion.
For three fields, these are
\begin{equation}
\begin{aligned}
\cM_{\sigma\sigma} &= \Omega^2 -\frac{1}{4} \eta \left(6 - 2\epsilon + \eta + 2 \xi \right)\\
\cM_{\sigma s} &= \Omega (-3 + \epsilon -\eta -\nu)\\
\cM_{\sigma b} &= -\Omega \tau.
\label{eq:kinetic_masses}
\end{aligned}
\end{equation}
\subsection{Dynamics}
Instead of describing the brief rapid turn with a potential and field space metric, following \cite{Fumagalli:2021mpc, Aragam:2023adu} we study a synthetic background evolution. We directly parameterize the time evolution of the turning rate $\Omega$ and torsion $\tau$ as
\begin{equation}
\begin{aligned}
    T(N_e) &= \left[\theta(N_e - (N_f - \delta/2)) - \theta(N_e - (N_f + \delta/2)) \right]\\
\end{aligned}
\label{eq:turn_profile}
\end{equation}
where $\Omega(N_e) = \Omega_0 T(N_e)$ and $\tau(N_e) = \tau_0 T(N_e)$, $N_e$ counts the e-folds after the beginning of inflation, and $N_f$ is the e-fold number at the center of the feature. For analytic convenience, we assume the turn is a top hat centered at time $N_f$ e-folds after the beginning of inflation, with width $\delta$, and height either $\Omega_0$ or $\tau_0$. Considering coincident profiles in $\Omega$ and $\tau$ is convenient, but it is not the most general case. Because torsion is not well defined if $\Omega$ equals zero, the most general situation would be a profile in $\tau$ inside the profile in $\Omega$.

Without choosing a potential and a field metric, the remaining masses in (\ref{eq:kinetic_pert_eom}) are unknown. We parametrize them in terms of the turn rates as
\begin{equation}
\begin{aligned}
\cM_{ss} &= \xi_{ss} (\Omega^2 + \tau^2) + \mathcal{M}_{ss,0}\\
\cM_{sb} &= \xi_{sb} \tau \\
\cM_{bb} &= \xi_{bb} \tau^2  + \mathcal{M}_{bb,0},
\end{aligned}
\label{eq:nonkinetic_masses}
\end{equation}
where $\xi_{ss},\xi_{sb},\xi_{bb},\cM_{ss,0},\cM_{bb,0}$ are assumed to be arbitrary real constants. We are unaware of any model where this parametrization of the masses is precisely valid, but take it as a natural generalization of the two-field scenario, where models with $\cM_{ss}$ an exact multiple of $\Omega^2$ are known, e.g. \cite{Achucarro:2019pux}. It is also possible to build models with arbitrary mass parametrization through a superpotential method \cite{Palma:2020ejf,Christodoulidis:2023eiw}. In field spaces with sufficiently many isometries, \cite{Christodoulidis:2022vww} found that rapid-turn, slow-roll trajectories must have $\cM_{sb}\sim (3-\epsilon)\tau$. 
Because this is of order of the friction terms we have already dropped, we choose $\xi_{sb} = 0 $ in all of our analytic calculations.

Although not an exact or approximate description of any concrete model's dynamics to our knowledge, the top-hat feature we study does qualitatively reflect many concrete models.
For example, the two-field models in \cite{Anguelova:2020nzl,Braglia:2020taf,Bhattacharya:2022fze} are concrete microphysics realizations of sharp inflationary features and generate gravitational wave spectra similar to the ones studied here.
In figure \ref{fig:turn_profile_Pz}, we show numerical results showing how the softening of the turn profile affects the primordial powerspectrum. Decreasing the sharpness of the feature but preserving the area has the effect of slightly reducing the enhancement in the high-$k$ tail of the feature, but otherwise gives a very similar $P_\zeta$ profile.
We discuss several ways these features might be observable in Section \ref{sec:Phenomenology}.

\begin{figure}[h]
\centering
\include{turnfigure}
\caption{The turn profile considered in this work. We show an exaggerated path in three-dimensional field space (left) and the turn rates as a function of time (right). On the left, the color changes from blue to red as a function of time. We label for future convenience three regions: I, II, and III, corresponding to before, during, and after the turn respectively. Note that the turn displayed on the left has a duration of $\delta = 1.1$ efolds for illustrative purposes, while values considered below consider much briefer turns, $\delta \leq 0.25$.}
\label{fig:turn_profile}
\end{figure}

\begin{figure}[h]
\centering
\includegraphics[width=0.9\textwidth]{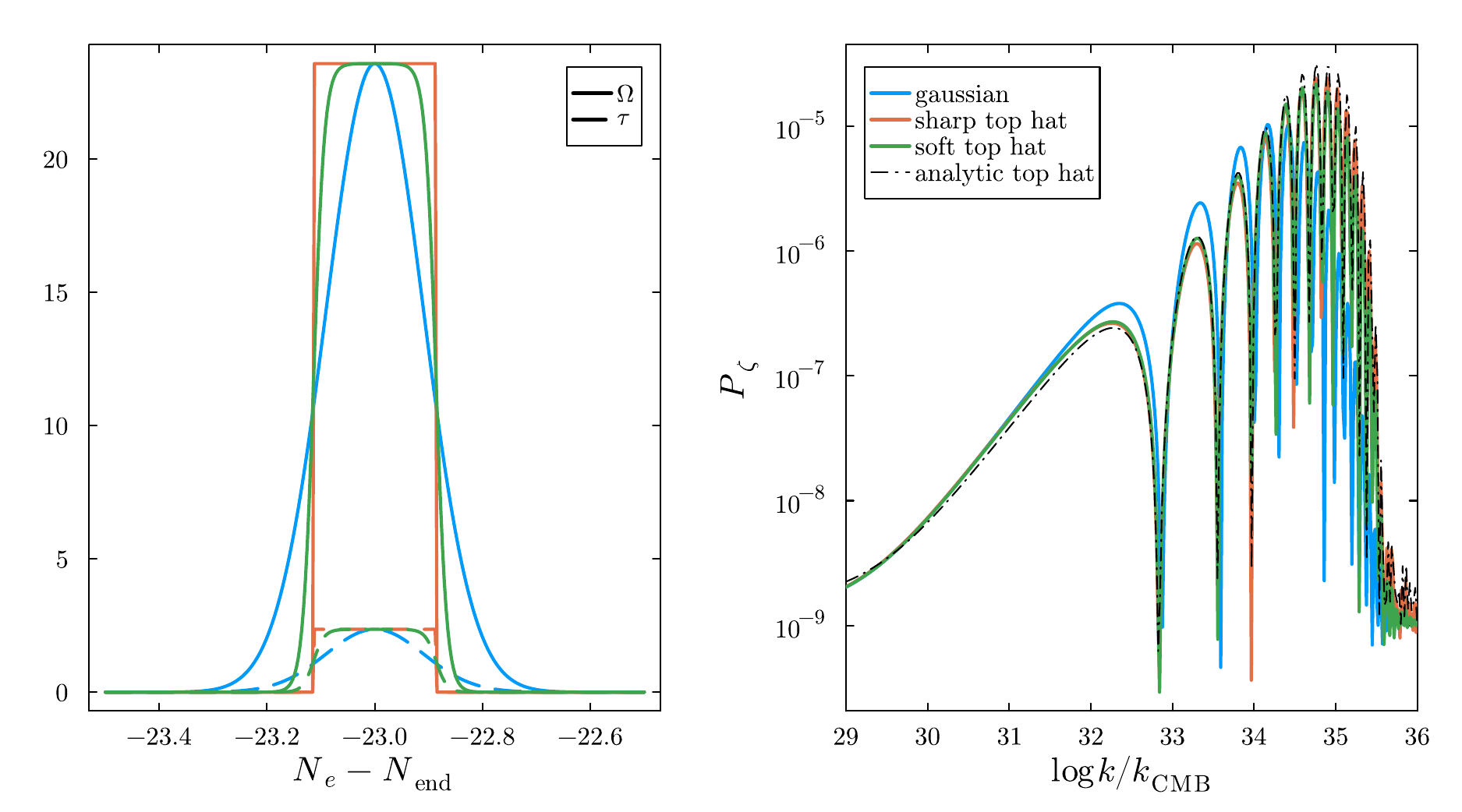}
\caption{We compare the power spectra (right) generated by different turn profiles (left) in fully numerical simulations using \texttt{Inflation.jl} \cite{Inflationjl}. Each run has $\rho = \Omega/\tau = 10$ and $\Omega^2+\tau^2 = 23.7^2$ and identical masses to the $\rho=10$ curves in Figure \ref{fig:inflationary-sgwb}. Instead of the analytic top hat profile \eqref{eq:turn_profile}, we take $T(N_e) =  \frac{1}{2}\left[ \tanh{\left(\frac{N_e+\delta/2}{N_b}\right)} - \tanh{\left(\frac{N_e-\delta/2}{N_b}\right)} \right]$, where we shift $T(N_e)$ so that the feature is centered 23 e-folds before the end of inflation. The ``sharp'' and ``soft'' top hats correspond to $N_b = \{10^{-4},10^{-2}\}$ respectively. The soft top hat is what was used in the numerical results of \cite{Aragam:2023adu}. The gaussian turn profile has $T(N_e) = \exp\left[-\left(\frac{N_e}{2\delta_g}\right)^2\right]$, with $\delta_g \equiv \sqrt{2\pi} \delta$ to give the top hat and gaussian cases equal integrated areas. Note also that the smoothed top hat has identical area to the analytic top hat for all values of $N_b$.
For all turn profiles, we see a qualitatively similar enhancement in $P_\zeta$, with the sharpest profile and the analytic result showing a $\sim 2$ times higher peak than the gaussian case at the large-$k$ end of the feature.
}
\label{fig:turn_profile_Pz}
\end{figure}

The brief turn in the background field trajectory causes perturbations to change from a quasi-single field Bunch-Davies initial state before the turn (region I) into an excited quasi-single field state after the turn (region III). The simple form of (\ref{eq:turn_profile}) allows us to use the WKB approximation to describe analytically the behavior of the perturbations during the turn (region II). The background metric and its first-time derivative are continuous at the junctions between regions I, II, and III. The corresponding matching conditions for the Mukhanov-Sasaki variables were derived by Deruelle and Mukhanov \cite{Deruelle:1995kd} and are
\begin{equation}
\begin{aligned}
\Delta(Q_{i}) &= 0\\
\Delta( \cD_N Q_{\sigma} - 2 Q_{s} \Omega ) &= 0\\
 \Delta(\cD_N Q_{s} -  Q_{b} \tau) &= 0\\
 \Delta(\cD_N Q_{b} + Q_{s} \tau ) &= 0,
\end{aligned} \label{matching-conditions}
\end{equation}
where the $\Delta$ operator matches quantities from region A to those from region B at junction time $t$ as $\Delta x \equiv x_A|_{t_{+}} - x_B|_{t_{-}}$.
These matching conditions are valid for any set of turn profiles $\Omega(N)$ and $\tau(N)$, even ones with a derivative discontinuity.

The perturbations in regions I and III are relatively simple to write down since they are quasi-single field. We describe region I in terms of a Bunch-Davies state:

\begin{align}
Q_{i,\mathbf{I}} &= u(\vec{k},N) \, \hat{a}_i + \mathrm{h.c.}(-\vec{k}) \\ 
u(k,N) &\equiv \left(\frac{i H}{\sqrt{2 k^3}}\right) \left(1- i \frac{k}{H a(N)}\right)e^{i k/ (H a(N))}  \hspace{5ex} &\text{for massless fields}\\ 
u(k,N) &\equiv \left(\frac{i H}{\sqrt{2 k^3}}\right) \left(\frac{- i e^{i (2\tilde{\nu}+1)\pi/4} \sqrt{\pi}}{\sqrt{2}} \right)\left(\frac{k}{H a(N)} \right)^{3/2} H_{\tilde{\nu}}^{(1)}\left(\frac{k}{H a(N)} \right)   \hspace{5ex} &\text{for massive  fields}
 \label{eq: conventions}.
\end{align} $\tilde{\nu}^2 \equiv \frac{9}{4} - \cM$, and $H_{\tilde{\nu}}^{(1)}$ are Hankel functions of the first kind ($\tilde{\nu} \equiv \sqrt{\tilde{\nu}^2}$ if $\tilde{\nu}^2>0$ and $\tilde{\nu} \equiv i \sqrt{-\tilde{\nu}^2}$ if $\tilde{\nu}^2<0$.)   Similarly, we describe region III in terms of an excited Bunch-Davies vacuum:
\begin{align}
Q_{i,\mathbf{III}} =(\alpha_{ij} u(k,N) + \beta_{ij} u^*(k,N)) \, \hat{a}_j + \mathrm{h.c.}(-\vec{k})
\label{eq: regionIII}
\end{align}

Note that our choice of including explicit constant contributions to the isocurvature masses is a first in the literature.
As we later describe in Section \ref{sec: Results}, depending on the nature of the feature it can source large growths in the isocurvature power spectra.
Without constant isocurvature masses, this isocurvature power would not decay, and any sharp feature would greatly enhance the adiabatic modes in its superhorizon. To maintain the known CMB amplitude, we would have to lower the scale of inflation very significantly, and would risk disturbing big bang nucleosynthesis and would remove any chance of sourcing a detectable SGWB.
In fact the presence of these masses was already taken as a default assumption in the literature, including in \cite{Palma:2020ejf,Fumagalli:2020nvq} and they were present during our numerical solutions of the perturbations' equations of motion in our previous work \cite{Aragam:2023adu}.

Imposing the canonical commutation relations for the fields and momenta is equivalent to requiring \cite{Fumagalli:2021mpc}: 
\begin{equation}
\begin{aligned}
\alpha_{ik}\alpha^*_{jk} - \beta^*_{ik}\beta_{jk} &= \delta_{ij} \\
\alpha_{ik}\beta^*_{jk} - \beta^*_{ik}\alpha_{jk} &= 0,
\end{aligned}
\label{eq:consistencyrelations}
\end{equation}
for each pair of fields $i,j$.

To describe the $\alpha_{ij}$ and $\beta_{ij}$ in terms of the initial state, we need to use the nontrivial matching procedure described above and a method to evaluate the perturbations in Region II. As noted above, we use the WKB approximation. Additionally, to simplify computations later on, we assume that the turn duration is so short that the effects of Hubble expansion on the perturbations can be ignored. In practice, we neglect all remaining terms proportional to $(3-\epsilon)$ in $F^a_b$ and $C^a_b$, which is equivalent to deriving the equations of motion from the Lagrangian \eqref{eq:pert_lagrangian} and disregarding any time derivatives of $a(t)$. In principle, this leaves $C^s_b = \cM^s_b$ as the only remaining nonzero off-diagonal element of $C^a_b$. However, we also neglect this term since $\cM_{sb}\sim (3-\epsilon)\tau$ in a wide range of models \cite{Christodoulidis:2022vww}.

To find the WKB approximation of the perturbations during the turn, we take the mode functions to have the form \cite{Bjorkmo:2019qno, Palma:2020ejf, Fumagalli:2020nvq} 
\begin{align}
Q_{i} = Q_{i,0} e^{i q N_e} \label{eq:Qtimeevol}
\end{align}
where the $Q_{i,0}$ are different for each field, and $q$ labels all of the WKB frequencies.
Plugging this into the frictionless equations of motion, we find that the $Q_{i,0}$ must be inter-related by
\begin{equation}
\begin{aligned}
Q_{s,0} &= i \frac{q^2 - \frac{k^2}{k_f^2}}{2 q \Omega_0} Q_{\sigma,0} \\
Q_{b,0} &= \frac{\tau_0}{\Omega_0} \frac{q^2 - \frac{k^2}{k_f^2}}{ \frac{k^2}{k_f^2}-q^2 + (\xi_{bb} - 1)\tau_0^2} Q_{\sigma,0}
\end{aligned},
\label{eq:QsRelation}
\end{equation}
and that there are six possible solutions $q = \pm \sqrt{\alpha_i}$, where the $\alpha_i$ are the three roots of the cubic polynomial. See \cite{Aragam:2023adu} for details.
In general, any solution in region II is a linear superposition of all six possible WKB exponents:
\begin{align}
Q_{i,\mathbf{II}} &= \sum_{j\in\text{fields}} \hat{a}_j \sum_{k=1}^{3}\sum_{\pm}Q_{ijk\pm,0} e^{\pm i \sqrt{\alpha_k} N_e} + \mathrm{h.c.}(-\vec{k})
\label{eq:regIIsolution}
\end{align}
The ($2 \cN^3=54$) $Q_{ijk\pm,0}$ amplitudes are independent along the solution axis ($k,\pm$), but they are inter-related via \eqref{eq:QsRelation} along the field labels $i$, resulting in a total of 18 independent amplitudes.

The expressions for the nine $\alpha_{ij}$ and the nine $\beta_{ij}$ in (\ref{eq: regionIII}) derived from the matching are very cumbersome and not immediately enlightening, so we have chosen to omit them from the text. However we have released them as supplementary code available with this paper, in both Mathematica and Julia, available at \url{https://github.com/rjrosati/3field-sharp-feature}.
In the uploaded notebooks, we include all $\alpha_{ij}, \beta_{ij}$, the WKB roots $\alpha_i$, a verification of the identities \eqref{eq:consistencyrelations}, and plotting code to reproduce almost all figures in this paper. Our code for performing the integral \eqref{eq:tensor_power_spectrum} will be released at a later date.

\subsection{The Tensor Power Spectrum}
\label{subsect: The tensor power spectrum}
The quantity of interest $\Omega_{\rm GW}$ is proportional to the tensor power spectrum \cite{Fumagalli:2021mpc},

\begin{equation}
h^2 \Omega_{\rm GW} = r_i \, {\cal P}_t (k, \tau_{\rm{end}})\equiv r_i \, {\cal P}_t (k),\hspace{4ex}  r_i \equiv h^2 \, 0.0416 \, c_g \,\Omega_{r,0} 
\end{equation} where $c_g \simeq 0.4$ and $\Omega_{r,0}$ is the energy density fraction in radiation today.
In the presence of excited states, and when non-Gaussian corrections can be neglected \cite{Garcia-Saenz:2022tzu}, 
\begin{equation}
\begin{aligned}
{\cal P}_t (k) &= \frac{H^4}{ 8 \pi^4 \Mpl^4 } \, \int_0^{\infty} dy \int_{|1-y|}^{1+y} dx \, \mu(x,y) \times  \\
&  \sum_{i,j} \left|\sum_{X, s_{1,2}=\pm} \alpha^{s_1}_{Xi} (x k) \alpha^{s_2}_{Xj} (y k) \cG_{XX}(s_1 x, s_2 y, z_{out})\right|^2 
\end{aligned}
\label{eq:tensor_power_spectrum}
\end{equation}
where, following the notation in \cite{Fumagalli:2021mpc}, we defined ${\alpha^{+}}_{ij}\equiv {\alpha}_{ij}$ and ${\alpha^{-}}_{ij}\equiv \beta_{ij}$. The geometrical factor is

\begin{equation}
\mu(x,y)=\left(\frac{(4 x^2- (1 + x^2-y^2)^2)}{4 x y }\right)^2
\end{equation} $\mu(x,y)$ varies between 0 and 1 and tends to zero as $x$ approaches $|1-y|$ and $1+y$. 
The equation (\ref{eq:tensor_power_spectrum}) generalizes equation (4.31) in the reference \cite{Fumagalli:2021mpc} slightly to include non-vanishing zero-masses $\cM_{ss,0}$ and $\cM_{bb,0}$.
\begin{equation}
\cG_{XX}(x,y,z_{\rm{out}}) \equiv \int_{z_{\rm{out}}}^{0} \frac{dz}{z^2} \, \zeta_X (x z) \zeta_X (y z)\, \frac{\zeta_{\sigma}(z) - \zeta^*_{\sigma}(z)}{2 i}
\end{equation}
where 
\begin{equation}
\begin{aligned}
\zeta_{\sigma}(z) & =  (1 + i z) e^{- i z} \\
\zeta_{s}(z) & =-e^{i(\nu_s + 1/2) (\pi/2)} \, i \sqrt{\frac{\pi}{2}}\, (- z)^{3/2} H^{(1)}_{\nu_s} (-z) \\
\zeta_{b}(z) & =-e^{i(\nu_b + 1/2) (\pi/2)} \, i \sqrt{\frac{\pi}{2}} \,(- z)^{3/2} H^{(1)}_{\nu_b} (-z) \\
\end{aligned}
\label{eq:mode_functions}
\end{equation}
\subsubsection{The Time Integral $\cG_{XX}$}
The time integral $\cG_{XX}$ can be computed exactly for some masses. When both fields are massless, the complete expression was given in section 4.4 of \cite{Fumagalli:2021mpc}. 

\begin{equation}
    \cG_{\sigma \sigma}(x,y,z)= 
    {\cK}(x,y)- {\cF}(x,y,z)- {\cF}^*(-x,-y,z)\\
\end{equation} where

 \begin{align}
   {\cK}(x,y) & = \frac{1- 2 x y - (x+y)^2}{(1- (x+y)^2)^2} \\ \nonumber \\
   {\cF}(x,y,z) & = \frac{e^{-i ( 1+x+y)z}}{2 ( 1+ x+y)^2} \times \\ \nonumber
    & \left(i \left[\frac{(1+x+y)^2}{z} - x y (1+x+y) z \right] -[x+y+(x+y)^2+ x y (2 + x + y)]\right) 
 \end{align}   
The only other case we know of with an exact closed form is when the dimensionless constant mass $\cM_{ss,0}=2$ and $\cM_{bb,0}=2$ . In that case

\begin{equation}
    \cG_{ss}(x,y,z)= 
    \tilde{\cK}(x,y)- \tilde{\cF}(x,y,z)- \tilde{\cF}^*(-x,-y,z)\\
\end{equation}where
 \begin{align}
  \tilde{\cK}(x,y) & = \frac{-2 x y }{(1- (x+y)^2)^2} \\\nonumber \\
   \tilde{\cF}(x,y,z) & = \frac{- e^{-i (1+x+y)z}}{2 ( 1+ x+y)^2} 
   \left(i \, x y \,(1+x+y)\,  z   + x y \, ( 2 + x + y)\right) 
 \end{align}   
As expected, $\cG_{\sigma \sigma}$ and $\cG_{ss}$ agree for large values of $x$ and $y$ where the mass becomes irrelevant, but they have different limits for small values of $y$ ($x$ is constrained to be in the interval $|1-y| \leq x \leq 1+y $). In the $y \ll 1$ limit

\begin{eqnarray}
    \cG_{{\sigma \sigma}} &=& \frac{3 z + 2 i - 2 i z^2}{4 z} + e^{- 2 i z} \left(\frac{- 2 i + z}{ 4 z} \right) + \cO (y, (x-1)) \\
     \cG_{{ss}} &=& y \left(  \frac{-3  + 4 i z + 2 z^2}{8} + e^{- 2 i z} \left(\frac{3+ 2 i  z}{ 8} \right) \right)  + \cO (y^2, y(x-1))\\
\end{eqnarray}
Regardless of the mass, the time integral has the following properties:

\begin{eqnarray}
    \cG_{XX}(x,y,z) &=& \cG_{XX}(y,x,z) \\ \nonumber
      \cG_{XX}(-x,-y,z)&=& \cG^*_{XX}(x,y,z) \\\nonumber
      \cG_{XX}(x,-y,z)&=& \cG^*_{XX}(-x,y,z) 
\end{eqnarray}

\section{Results}
\label{sec: Results}

\begin{figure}[t]
    \centering
    \includegraphics[width=\textwidth]{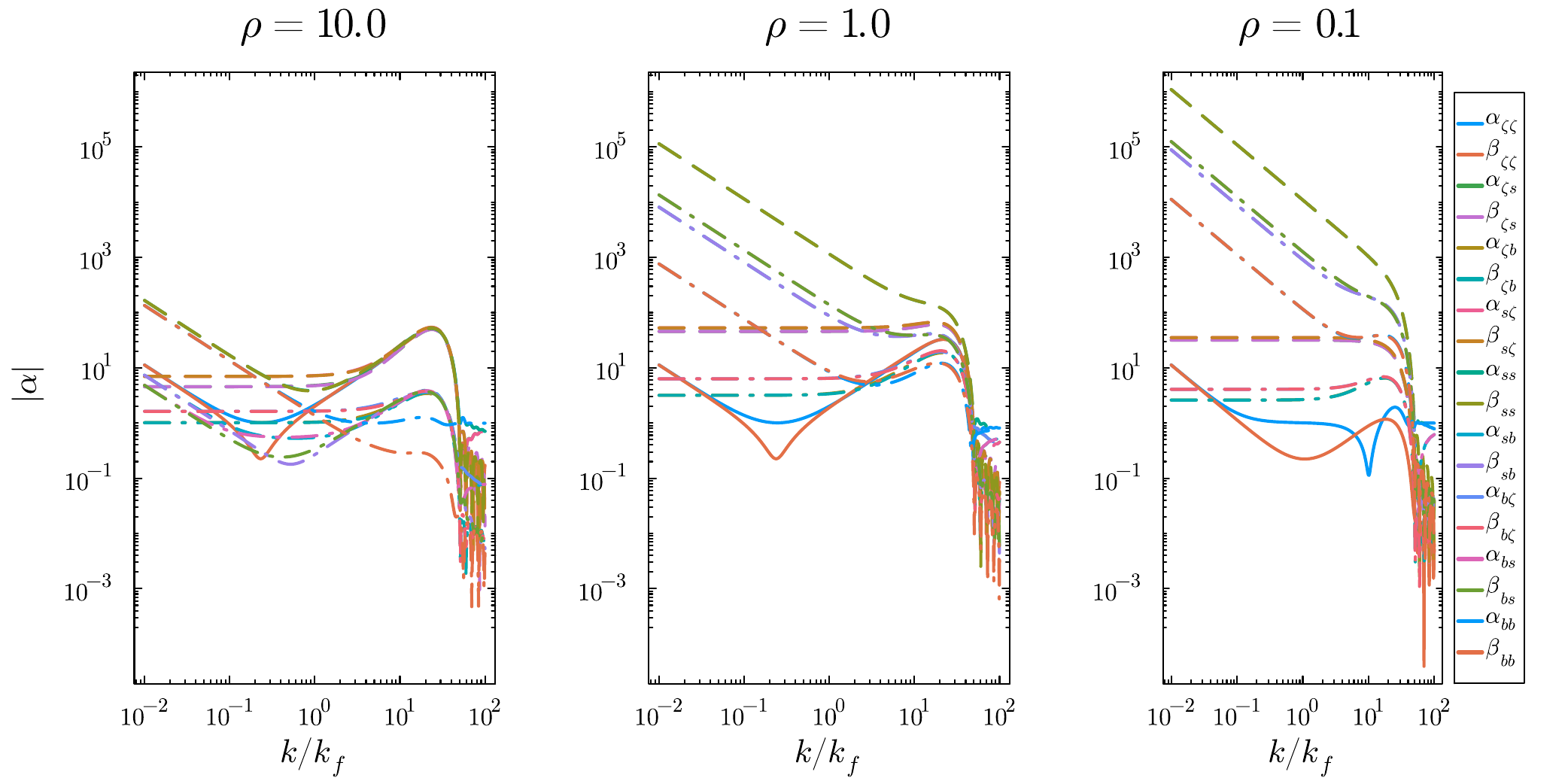}
    \caption{A comparison of the Bogoliubov coefficients as a function of $\kappa$ for three values of $\rho \equiv \Omega/\tau$ and a fixed $\Omega_{2f} = \sqrt{\Omega^2 + \tau^2} = 23.7$ and $\delta=0.225$. The masses are parameterized by \eqref{eq:nonkinetic_masses} with $\xi_{ss}=-3, \xi_{bb}=2, \cM_{ss,0} = \cM_{bb,0}=2$. Solid lines represent $\alpha_{\zeta \zeta}$ and $\beta_{\zeta \zeta}$. Dashed lines correspond to Bogoliubov coefficients for which at least one of the indices is an “s.” Dot-dashed lines correspond to Bogoliubov coefficients with at least one index being a “b.”
    The quasi-two-field, $\Omega$-dominated case on the left has all coefficients contributing approximately equally and a sharp peak around $k \sim \Omega_{2f} k_f$, while a strong hierarchy develops as the torsion increases. The dominant coefficients rapidly become $\alpha_{ss}, \beta_{ss}$ and the peak vanishes, replaced by a steep powerlaw growth towards the superhorizon.}
    \label{fig:bog_comparison}
\end{figure}
\begin{figure}[t]
    \centering
    \includegraphics[width=\textwidth]{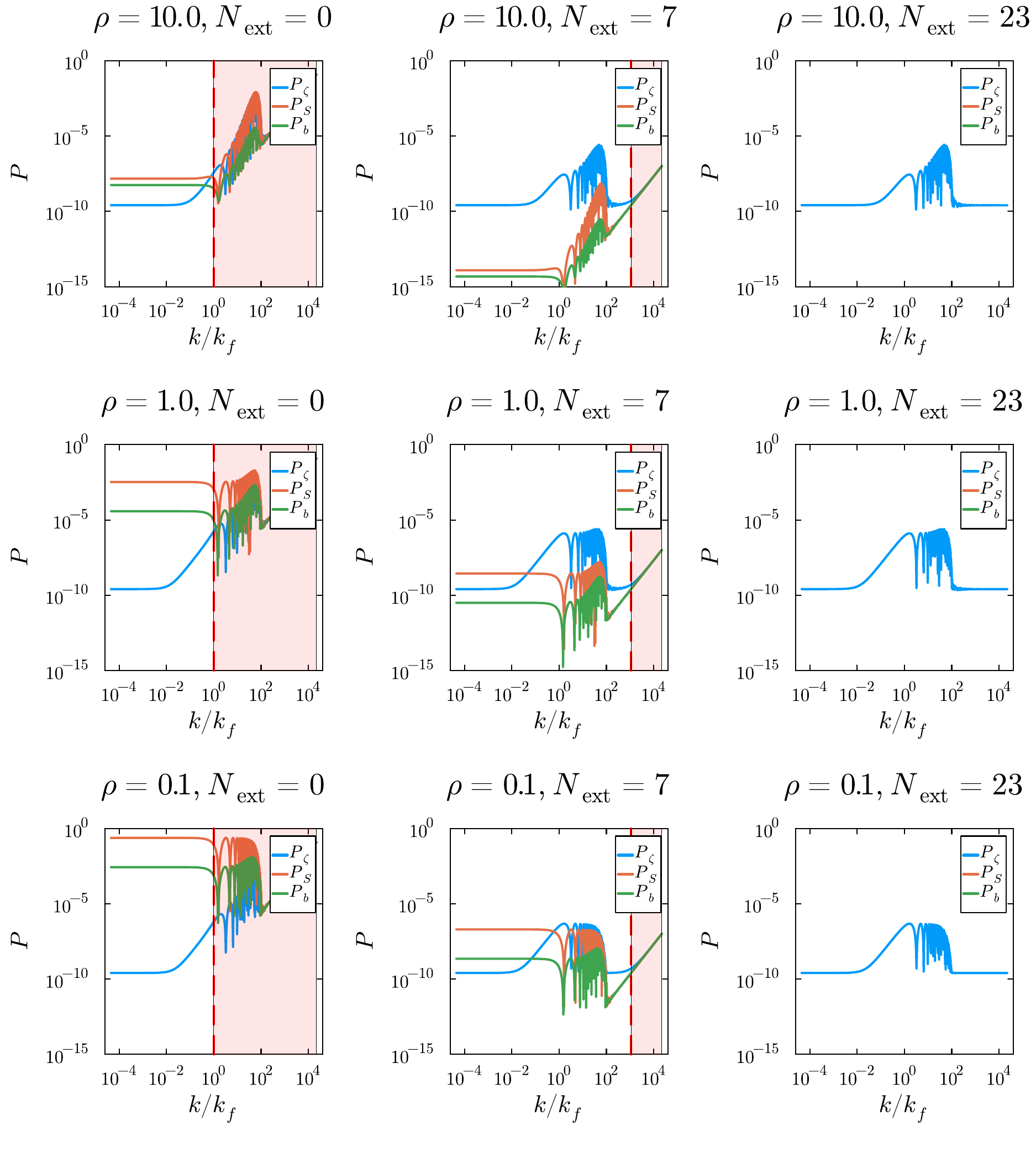}
    \caption{We show the approximate power spectra $P_i(k,N) = \zeta_i(k,N)^2 \sum_{j} |\alpha_{ij}(k) + \beta_{ij}(k)|^2$ at the time of the feature (left), 10 e-folds after (middle), and 23 e-folds after (right) for three values of the turn rate ratio $\rho$ (rows). We plot the out-region estimates of the power spectrum shape, but scale them with the appropriate time dependence of the mode functions \eqref{eq:mode_functions}. These approximate power spectra are therefore a poor approximation for the modes that have not yet exited the horizon at the plotted time, regions we have shaded in red in these plots. The isocurvature power spectra decay rapidly outside the horizon. This plot corresponds to $\Omega_{2f}=50.0, \xi_{ss}=-3,\xi_{bb}=2,\delta=0.1$. }
    \label{fig:powerspectra_decay}
\end{figure}

In this section we investigate the structure of the three-field excited state after the sharp feature, as well as the inflationary-era graviatational stochastic background it sources.

In Ref. \cite{Fumagalli:2021mpc}, the scaling of the inflationary-era SGWB amplitude was predicted to go with the number of fields $\cN$ as $\cN^4$, under the assumption that the Bogoliubov coefficients' shape in $k$ remains roughly constant and the additional fields enter at the same amplitude.
We examine this conjecture in Fig. \ref{fig:bog_comparison}, which plots the values of the $2\cN^2=18$ Bogoliubov coefficients for three representative values of $\rho=\Omega/\tau$ and a fixed $\sqrt{\Omega^2+\tau^2}=23.7$.

\begin{figure}[t]
    \centering
    \includegraphics[width=0.98\textwidth]{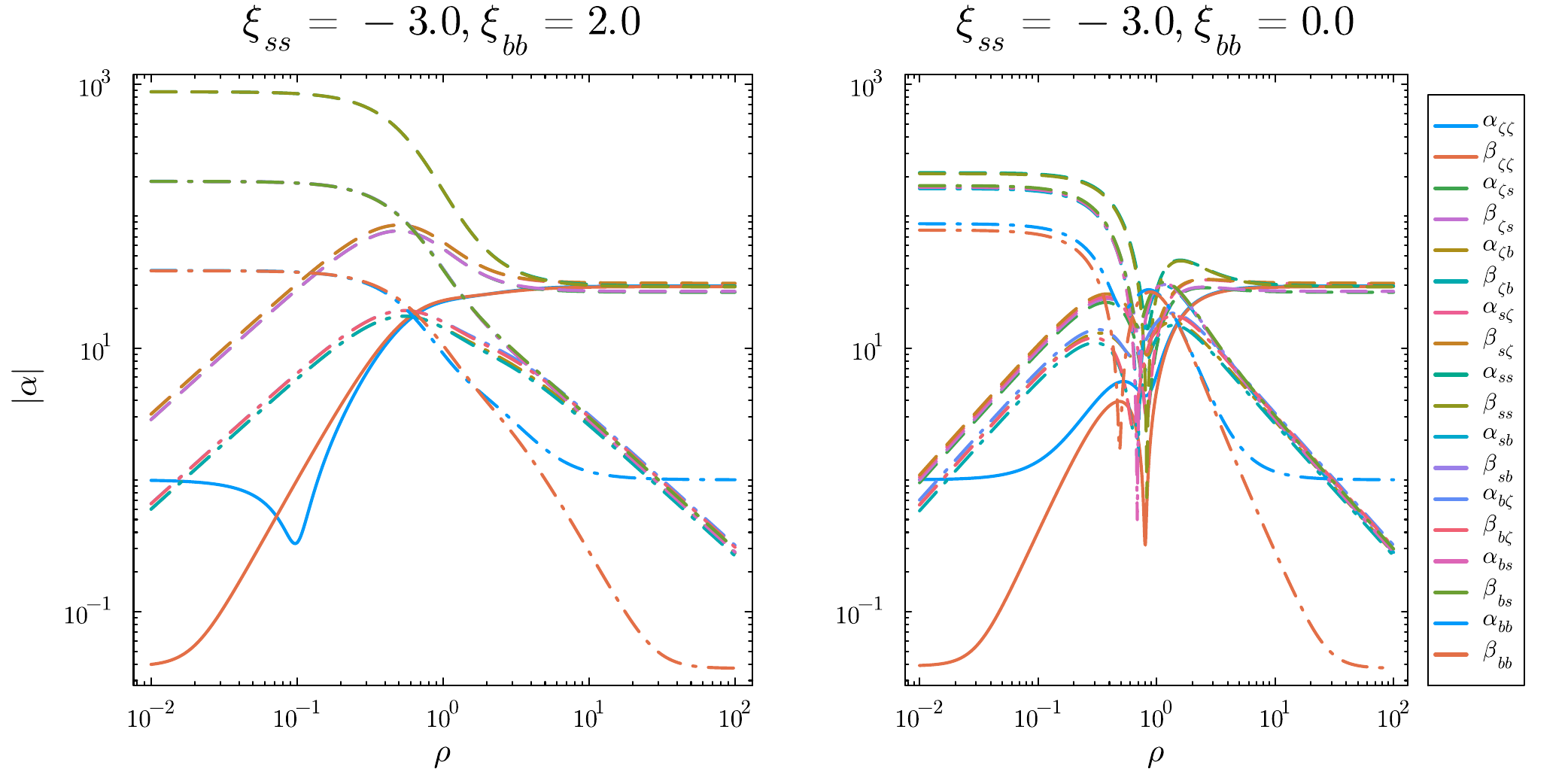}
    \caption{This plot shows how the Bogoliubov coefficients vary as a function of $\rho=\Omega_0/\tau_0$ in the interval of values used to plot Fig \ref{fig:bog_comparison}, fixing $\kappa=\Omega_{2f}/2, \xi_{ss}=-3, \Omega_{2f} = 23.7, \delta = 0.225$. We compare the variety of possible growths with two values of $\xi_{bb}$. At high $\rho$ (two-field limit), we see the ($\zeta$, $s$) 2$\times$2 block of Bogoliubov coefficients reaches approximately the same value. At lower $\rho$, the coefficients split into a strong hierarchy, with the relative amplitudes set by $\xi_{bb}$ and the $ss$-coefficients the highest amplitude. }
    \label{fig:ass(rho)}
\end{figure}

At superhorizon scales, the rapid turning causes both adiabatic and isocurvature modes to be activated, with the latter being more prominent, as seen in Figure \ref{fig:bog_comparison}. In order to match the Planck bounds on isocurvature power, the growth is restricted by giving the isocurvature modes a non-zero mass. For $\cM_{ss,0}=\cM_{bb,0}=2$, the Bogoliubov coefficients growth is proportional to $k^{-1}$ (for smaller masses, $\alpha_{ss} \sim k^{-g}$ with $g>1$), and the amplitude depends heavily on $\rho$, as shown in Figure \ref{fig:ass(rho)}. The $k^{-1}$ momentum-dependent behavior leads to the isocurvature power spectrum being flat for low $k$ because the massive wave functions (\ref{eq: conventions}) behave as $ 1/ \sqrt{k}$ when $k \ll k_f$. As the modes move outside the horizon, the generated isocurvature power decreases rapidly. Figure \ref{fig:powerspectra_decay} illustrates the behavior of the power spectra at three different time points, which is consistent with current constraints on isocurvature modes.

   The left plot in Fig. \ref{fig:bog_comparison} corresponds to $\Omega \gg \tau$ and matches the two-field results of Fig. 3 in \cite{Fumagalli:2021mpc} in the region the plots overlap. Initial states with momenta larger than $2 k_* = 2  \Omega_{2f} k_f$ are not excited and remain in the Bunch Davies vacuum. Below this momentum cutoff, there are always excited states, but how much each field is excited depends on the value of the momenta. In this regime, $\{\alpha_{\zeta \zeta},\beta_{\zeta \zeta}, \alpha_{s \zeta},\beta_{s \zeta}, \alpha_ {\zeta s}, \beta_{\zeta s},\alpha_{ss}, \beta_{ss} \}$ are equally dominant around $k_*$. This gives the $2^4$ enhancement with respect to the single field case found in \cite{Fumagalli:2021mpc} for the two-field case, but the factor of $2^4$ does not translate into a $\cN^4$ behavior for $\cN$ fields. Because torsion is subdominant to the turning rate, Bogoliubov coefficients with an index ``b" are subdominant, and hardly contribute to sourcing the SGWB (\ref{eq:tensor_power_spectrum}). 

The middle plot in  Fig. \ref{fig:bog_comparison} corresponds to  $\Omega \sim \tau$. As in the previous case, modes with $k> 2 k_*$ remain in the initial Bunch Davies vacuum. While there is still a peak around $k_*$, the contributions from the different Bogoliubov coefficients develop a hierarchy. The largest contribution corresponds to $\{\alpha_{ss}, \beta_{ss} \}$, followed by the contribution from $\{\alpha_{\zeta s},\beta_{\zeta s }, \alpha_{ s \zeta},\beta_{s \zeta} \}$ which is down by a factor of 2, followed by $\{ \alpha_{b s},\beta_{b s }, \alpha_{ s b},\beta_{s b},\alpha_{zz}, \beta_{zz} \}$ down by approximately another factor of 2. Though more Bogoliubov coefficients are activated, there is a hierarchy and the power spectrum can't grow like $\cN^4$. Overall, the maximum value of $\cP_t$ is larger for  $\Omega\sim \tau$ than for $\Omega\gg \tau$ when keeping $\sqrt{\Omega^2+\tau^2}$ fixed.  At superhorizon scales, the growth is dominated by $\{\alpha_{ss}, \beta_{ss} \}$. These two coefficients have the same absolute value as Fig. \ref{fig:bog_comparison} shows and is expected from (\ref{eq:consistencyrelations}). 
 

 The plot on the right in  Fig. \ref{fig:bog_comparison} corresponds to  $\Omega \ll \tau$. In this regime, the peak around $k_*$ has nearly disappeared. The Bogoliubov coefficients $\{\alpha_{ss}, \beta_{ss} \}$ dominate over all others, but with the same small momenta dependence,  $k^{-1}$. In this limit $|\alpha_{\zeta X}|$ and $|\beta_{\zeta X}|$ are several orders of magnitude smaller than  $\{|\alpha_{ss}|, |\beta_{ss} |\}$, which is consistent with the observation in \cite{Christodoulidis:2022vww} that on superhorizon scales the evolution of the adiabatic modes and the isocurvature modes decouple.

In the previous paragraph, the results may not be consistent with first-order perturbation theory. In Figure \ref{fig:powerspectra_decay}, we have plotted the power spectrum for both adiabatic and isocurvature modes at three different times: when the feature is produced ($N_{\rm{extra}}=0$), ten e-folds after the feature ($N_{\rm{extra}}=10$), and twenty-three e-folds after the feature ($N_{\rm{extra}}=23$), for three different values of the ratio $\rho=\Omega/\tau$. The power spectra at late times match experimental data in all cases, but for $\rho=0.1$, the isocurvature power spectrum at ($N_{\rm{extra}}=0$) briefly reaches $\mathcal{O}(1)$, calling into question the validity of first-order perturbation theory.


In the second part of the results, we discuss the shape of the fractional energy density spectrum ($\Omega_{GW}$). Scalar fluctuations source gravitational waves at two different times: during the radiation-dominated era as the perturbations re-enter the horizon and during inflation. In our previous work \cite{Aragam:2023adu}, we calculated the contribution from gravitational waves generated during the radiation-dominated era, which were caused by adiabatic scalar fluctuations reentering the horizon.  The frequency profile of $\Omega^{\rm{rad}}_{GW}$ changes substantially as a function of turning rate over torsion, $\rho$. When $\Omega \gg \tau$, $\Omega^{\rm{rad}}_{GW}$ peaks at $k_*$, while for $\Omega \ll \tau$ the peak has migrated below $k_f$. Said differently, the profile depends on the number of active fields.

On the other hand, the profile of $\Omega^{\rm{inf}}_{GW}$ is not greatly affected by the relative ratio of $\Omega$ to $\tau$ or the effective kinetic masses during the feature.
In Figure \ref{fig:inflationary-sgwb}, we compare the relative amplitudes and shapes of $\Omega^{\rm{inf}}_{GW}$ as a function of $\rho$.
The background peaks at approximately $k_f$ and, for $k>k_f$, oscillates with a frequency $2/k_f$. This result is independent of the effective number of acting fields. The amplitude depends on the number of fields, but not simply as a power of the number of active fields ${\cal N}$. The actual dependence is more complicated. Figure \ref{fig:ass(rho)} shows how the dominant $\alpha_{ss}$ and $\beta_{ss}$ grow with the ratio of the turning rate over the torsion, ${\rho}$. This continuous variable interpolates between two active fields in the limit of vanishing torsion, and three active fields when the torsion dominates over the turning. The significant growth of the perturbations when ${\rho \ll 1}$ raises questions about the reliability of perturbation theory to compute $\Omega_{\rm{GW}}$ as we discuss in the next section.

\begin{figure}[t]
\centering
\includegraphics[width=\textwidth]{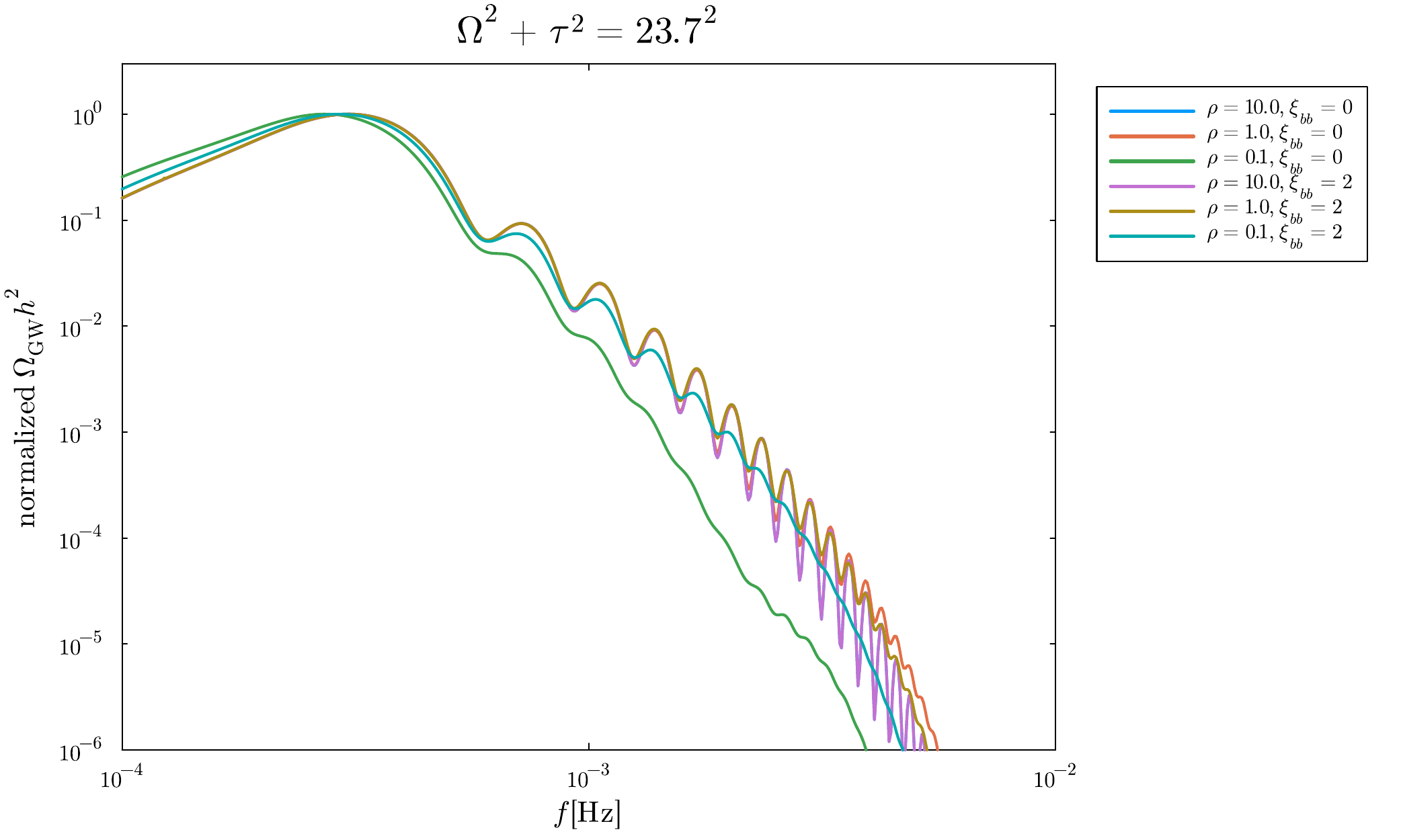}
\caption{We compare the possible shapes of $\Omega^{\rm{inf}}_{GW}$ from our sharp feature, normalizing by the maximum amplitude to focus on the shapes of the signals. These signals all occur at the same feature scale and share the parameters $\xi_{ss}=-3,\delta=0.225,\Omega_{2f}=23.7$, while we vary the turn rate ratio and the kinetic coupling of the third mass $\xi_{bb}$. The mass content and turn rate ratio $\rho$ do not strongly affect the shape of the signal in frequency, especially near the peak. The $\rho=0.1, \xi_{bb}=0$ case does show some deviation in the envelope of the signal towards the subhorizon of the feature (high $f$). We caution that the largest enhancements we present with $\rho=0.1$ are likely subject to backreaction corrections. The relative similarity of these signals despite the large variation in the structure of the corresponding excited states leads us to call this SGWB profile ``universal''.}
\label{fig:inflationary-sgwb-masses}
\end{figure}

In Figure \ref{fig:inflationary-sgwb-masses}, we compare the variety of inflationary SGWB shapes that we are able to produce in this work.
Despite the variations in $\rho$ and the mass parameters leading to fairly variable structures of the excited state (cf. Figures \ref{fig:bog_comparison}, \ref{fig:ass(rho)}), the inflationary-era SGWBs remain approximately the same shape, at least around the main peak.
As we discuss in Section \ref{sec:Phenomenology}, discerning these small differences would require resolving features of the signal at least $10^2$ or $10^3$ times lower amplitude than the peak, so except in the case of a very loud signal these shapes would be equivalent from a data analysis perspective.

\section{Backreaction}
\label{sec:Backreaction}
To detect signals at LISA, the scalar perturbations must be enhanced by at least $10^4$, as illustrated in Fig \ref{fig:powerspectra_decay}. Such a significant enhancement raises doubts about the reliability of perturbation theory and has been a concern of previous work on this subject \cite{Fumagalli:2020nvq, Fumagalli:2021mpc, Inomata:2022yte, Ota:2022xni,Fumagalli:2023loc}. There are two conditions for trusting perturbation theory \cite{Holman:2007na, Bartolo:2013exa, Cannone:2014qna, Adshead:2014sga}: (i) ensuring that the energy in the perturbations is much smaller than the energy in the background field not to disrupt the background equation of state and (ii) ensuring the reliability of the linear equations for the evolution of the perturbations.  Both concerns were addressed in \cite{Fumagalli:2020nvq, Fumagalli:2021mpc} for the two-field equivalent of the model considered in this work and \cite{Aragam:2023adu} for the three-field case.

It is easy to show that (i) is easily satisfied in the cases studied in this work. According to \cite{Holman:2007na}, the contribution of the perturbations to the energy density is:
\begin{equation}
    \rho_{\rm{excited} \,\,\rm{ states}}= \frac{1}{a^4} \sum_i \int \frac{d^3 k}{(2 \pi)^3} \, k\, \left( |\beta^{\zeta i}_k|^2 +|\beta^{si}_k|^2 +|\beta^{bi}_k|^2\right) 
\end{equation}
When $\Omega \lesssim\tau$, it is easy to compute this quantity. In the comparison shown in Fig \ref{fig:bog_comparison}, $\beta_{ss}$ dominates. For the mass choices made in this calculation, $|\beta_{ss}| \propto 1/k$ when $k< \Omega_{2f} \, k_f$, and it vanishes when $k> \Omega_{2f} \, k_f$.
\begin{equation}
    \rho_{\rm{excited}} \sim \frac{3}{a^4} k_f^4 = 3 H^4
\end{equation}
The constraint $ \rho_{\rm{excited}} \ll \rho_{\rm{inf}}$ is always satisfied as 
\begin{equation}
    \rho_{\rm{excited}} \sim  3 H^4 \ll 3 M^2_{\rm{Pl}} H^2
\end{equation} 
In the two-field case, \cite{Fumagalli:2021mpc} showed that the second constraint is always the most stringent. For two fields,  the backreaction constraint (i) equals the linearity constraint (ii) times $\epsilon$. The $\epsilon$ difference between the two constraints may also explain the outcome of the lattice computations carried out in a different two-field model \cite{Krajewski:2022uuh}. In this case,  the energy density in the perturbations is approximately $10^{-2} \rho_{\rm{inf}}$. Still, the simulation shows that perturbation theory breaks down as the perturbations don’t grow as much as expected in perturbation theory.  Their effect is to shift the background to a new classical configuration. Because in \cite{Krajewski:2022uuh} $\epsilon \lesssim 10^{-2}$, the analysis presented in \cite{Fumagalli:2020nvq, Fumagalli:2021mpc} leads to the expectation that constraint (ii) is not satisfied even if (i) is. Non-perturbative effects have been also shown to be important in a single-field inflation model with a departure from slow-roll \cite{Caravano:2024tlp}.

In the three-field case, the constraints analysis performed in \cite{Aragam:2023adu} only considered the enhancement in the adiabatic power spectrum as this was the only one to survive until radiation-dominated re-entry. But Fig. \ref{fig:powerspectra_decay} shows that the isocurvature power spectrum is several orders of magnitude larger at the time of the feature, for example, when $\rho=0.1$, $P_s\sim {\cal O}(1)$, and, hence, ${\cal O}(Q_s/M_{\rm{Pl}}) \sim 1$. This behavior points to a brief breakdown of perturbation theory because the cubic Lagrangian \cite{Pinol:2020kvw} contains terms such as
$$ {\cal L}_3 \supset a \,\sqrt{2 \epsilon} \,\Omega \,M_{\rm{Pl}}\, Q_s \, (\partial \zeta)^2$$ which is larger, for a brief time around the feature,  than the quadratic term $$ {\cal L}_2 \supset a \, \epsilon \, M_{\rm{Pl}} \,(\partial \zeta)^2 $$

We therefore caution that the largest enhancements we present in this work with short periods of $P_S\sim 1$ would experience strong corrections from backreaction effects in full lattice simulations\footnote{In addition to the tree level results presented here and in other work, a one-loop calculation was performed in \cite{Fumagalli:2023loc}. Although a full non-perturbative analysis is likely necessary to accurately predict the SGWB in the highest torsion cases we present.}.
The brief high amplitude of the isocurvature power spectra around the time of the feature would likely be dampened, leading to less power in gravitational waves.
Nonetheless, the results we present at lower enhancements remain valid and experience only small corrections from backreaction.

\FloatBarrier

\section{Phenomenology}
\label{sec:Phenomenology}

In this section we discuss the observable consequences of sharp features during inflation, and how they impact the possible cosmological observables at CMB scales, LSS, and of course a possible SGWB. We also comment on the detectability of the observable probes, and how likely a possible detection is to well-constrain the sourcing feature.

The toy model of sharp feature we study in this work generates a linear in $k$ oscillatory enhancement in the adiabatic power spectrum and a similar-magnitude temporary enhancement in the isocurvature power spectra, which decay to unobservability within a few e-folds when they have positive masses. When sufficiently strong, these features can generate radiation-era and inflationary-era SGWBs.
But observing the effects of a sharp feature in any particular cosmological probe is very dependent on when exactly it occurs during inflation, or equivalently the value of the scale of the feature $k_f$.
Although not the focus of this work, we will briefly comment on non-GW observables from inflationary sharp features.
Current constraints on the primordial adiabatic power spectrum rule out large enhancements at CMB scales, but perhaps sufficiently low-amplitude features could be visible at CMB scales as ``primordial clock'' oscillations in the adiabatic power spectrum  \cite{Chen:2011zf}.
The current constraints from LSS place many limits on the primordial adiabatic power spectrum at scales smaller than the CMB, including not disrupting baryogenesis $P_\zeta(k) \lesssim 10^{-2}$, and constraints from structure formation (see \cite{Slosar:2019gvt,Achucarro:2022qrl} for a recent review). Otherwise, LSS allows for large scalar perturbations at small scales, which can source the radiation-era SGWB we studied in part I of this work \cite{Aragam:2023adu}.
We note that these cosmological probes also constrain and limit isocurvature power -- the features we study do not predict any observable isocurvature, as the brief spike in isocurvature decays with a small positive mass long before undergoing reheating and contributing to either the CMB or LSS.
Non-gaussianities will be another constraining observable.  They can be significant in multifield models with sustained turning \cite{Garcia-Saenz:2018ifx, Garcia-Saenz:2018vqf, Fumagalli:2019noh, Iarygina:2023msy} and large-scale features \cite{Chen:2010xka}. A recent computation \cite{Iacconi:2023slv} found them to also be significant for small-scale large features.  Although the models in this analysis differ from those presented in this work, we expect their result to hold. 

Gravitational waves provide perhaps the most undistorted probe of the inflationary era, and the inflationary-era SGWB \textit{is uniquely sensitive} to the entire evolution of the perturbations, even transients that do not leave a signature in the perturbations at the end of inflation.
The signal we study in this work (cf. Figure \ref{fig:inflationary-sgwb}) fits a roughly broken powerlaw shape, with numerically measured powerlaws of $k^{3}$ on the low-$f$ tail and $k^{-3}$ to $k^{-4}$ fitting the first 3-4 peaks, and $k^{1.7}$ and $k^{-8}$ fitting just the most prominent peak.
The signal has $\cO( 10\%)$ oscillations linear in $k$ beginning at the peak and with frequency $\omega \sim 2/k_f$.
The shape of the signal is surprisingly largely independent of the mass parameters \eqref{eq:nonkinetic_masses} and even the torsion to turn rate ratio $\rho$ for fixed $\Omega_{2f}$ (cf. Figure \ref{fig:inflationary-sgwb-masses}). The amplitude, however, rises quite steeply as the torsion increases (see the growth of the Bogoliubov coefficients in Figure \ref{fig:ass(rho)} as $\rho \rightarrow 0$). We numerically measure $\Omega_{\rm GW} \propto \tau^{24}$ (!) in the regime of steepest growth around $\rho = 1$ for a fixed $\Omega_{2f}$ and $\xi_{bb}=2$, although we caution that the largest enhancements are almost certainly subject to backreaction corrections. The envelope of the inflationary-era SGWB at high $k$ does begin to change at sufficiently high $\tau$, but the peak remains unaffected.
Because the shape of the inflationary-era SGWB near its peak does not depend on the details of the feature (and as argued below, this generalizes to more realistic models), we call it a \textit{universal signature of sharp features in multi-field inflation}
\footnote{It is intriguing that various mechanisms for generating sharp features display a similar envelope for $\Omega^{\rm{inf}}_{\rm{GW}} (k)$ \cite{An:2020fff, An:2022cce}. Comparing the specific broken power law exponents around the prominent peak would be valuable to determine whether a potential detection could provide additional information beyond merely confirming the existence of a sharp feature.}.
Though a similar universal phenomenology was claimed in Ref. \cite{Fumagalli:2021mpc}, they assumed the Bogoliubov coefficients were strongly peaked and approximately equal magnitude.
We have found that the excited state takes on a significantly different form with a strong hierarchy in the coefficients (cf. Figure \ref{fig:bog_comparison}), and that the amplitude of the SGWB depends strongly on the entropic sector of the perturbations with growth that vastly outpaces the predicted $\cN^4$.
This enhancement is so strong that it opens a new window of observable parameter space: we may source a detectable inflationary-era SGWB without requiring a large enhancement in the adiabatic power spectrum.

If such a signal were detected, its shape would immediately reveal the feature scale from the oscillation frequency $2/k_f$, and confirm that an excited state occurred during inflation in the early universe.
Interpreting the signal in terms of multi-field dynamics would be difficult due to its somewhat universal nature, unless the radiation-era signal were also visible (cf. Figure \ref{fig:inflationary-sgwb}). Because the radiation-era feature is only sourced by the adiabatic power spectrum, the ratio of the two amplitudes gives information about the effective amplitude of the turn $\Omega_{2f}$ and the number of dynamically contributing fields.
Parameter estimation studies on the recovery of templated SGWBs with similar features to the one presented here are available in \cite{Fumagalli:2021dtd,Braglia:2024kpo}.
The equation of state of the universe could also modify the post-inflationary scalar-induced portion of the signal if different than radiation \cite{Witkowski:2021raz}.

Of course physical inflationary backgrounds are quite unlikely to generate features extremely similar to the top hat we study \eqref{eq:turn_profile}, though isolated single turns with a smoother profile can occur in some backgrounds \cite{Anguelova:2020nzl}.
One common source of sharp features in multi-field inflation occurs when the trajectory becomes unstable (e.g., due to a sufficiently tachyonic entropic mass)\footnote{Having a tachyonic mass is not the precise criterion for instability of the equations of motion \eqref{eq:kinetic_pert_eom}. For three-field inflation these criteria have been studied in \cite{Christodoulidis:2023eiw, Christodoulidis:2022vww}.} and the fields change direction abruptly.
Several classes of models in the literature fit this description including waterfall mechanisms \cite{Linde:1993cn} and geometric destabilization \cite{Renaux-Petel:2015mga}. Some of the present authors also saw similar features in a $\cO(100)$-field, random potential inflation model as the trajectory approached saddle points \cite{Paban:2018ole}.
Multi-field models often give rise to attractor behavior in field space and sharp features can even lie on the attractor, as seen in \cite{Geller:2022nkr,Lorenzoni:2024krn}.

References \cite{Braglia:2020taf, Bhattacharya:2022fze} studied several such two-field models that produce observable signals in LISA -- these features typically are not one top hat as we've studied here, but a series of sharp turns with decaying amplitude, although the observed power spectra and radiation-era SGWBs (inflationary-era SGWBs are not studied in their work) are qualitatively very similar to the two-field limit of ours.
Although we consider a full investigation outside the scope of this work, we have generalized some of the models in \cite{Braglia:2020taf} to three fields and have found that the features' behavior persists, and we see decaying oscillations in both turn rate and torsion during the trajectory's realignment. We show some of these dynamics in Appendix \ref{sec:bch}.
Reference \cite{Christodoulidis:2023eiw} has also studied three-field backgrounds with more broad features or constant turning and found large enhancements.

The features we study here, then, are by no means an exact quantitative prediction for a physical model's SGWB, but are qualitatively similar. It remains an open question how universal the shape we have found is to physical backgrounds with sharp features, but we suspect they will be qualitatively very similar.
The Green's functions and kernel in \eqref{eq:tensor_power_spectrum} are fixed by the physics and set the behavior of the background at low- and high-$k$. And despite a very different structure of excited state when $\tau \gg \Omega$ with a top hat, we find a similar shape of SGWB.

Similarly, we might speculate on how more dynamical fields might affect the SGWB.
We do not have a strong expectation for the structure of the excited state, but the above arguments about the Green's function and kernel still apply so we expect the qualitative nature of the SGWB to remain the same.
Because the entropic sector of the perturbations can experience a large enhancement during the feature and all fields contribute to the SGWB \eqref{eq:tensor_power_spectrum}, all else equal, we expect more fields to increase the amplitude of the inflationary-era SGWB compared to the radiation-era one, and allow for a LISA-detectable SGWB without large scalar perturbations.
The radiation SGWB can also be relatively suppressed depending on the duration of the feature \cite{Fumagalli:2021mpc}, so its absence is not a smoking gun for more dynamical fields.

\section{Conclusions}

In this work, we have completed the inflationary-era contribution of the stochastic gravitational wave background sourced by a sharp turn during three-field inflation, extending our previous work computing the radiation-era contribution in \cite{Aragam:2023adu}.
These two contributions together complete the calculation of the expected SGWB signal from a sharp feature up to tree-level in the perturbations.

From an phenomenological point of view, the most interesting results of this work are that: 1) because of temporarily large isocurvature power, sharp features in multi-field inflation have the ability to source large inflationary-era scalar-induced SGWBs ($\Omega_\mathrm{GW}^\mathrm{inf}$) without requiring large adiabatic enhancements and 2) sharp features in multi-field inflation make a somewhat universal shape of $\Omega_\mathrm{GW}^\mathrm{inf}$, c.f. Figure \ref{fig:inflationary-sgwb}.

We now elaborate on these points as well as describe the large-$\cN$-field limit.
The main differences between the two-field case studied in \cite{Fumagalli:2021mpc} and the three-field case reported here are the following:

\begin{itemize}

\item In the presence of torsion, it is necessary to include a constant term in the perturbations' entropic masses, as otherwise, the growth of the isocurvature perturbations outside the horizon would clash with the bounds on isocurvature power measured by Planck. We included these masses in our computation of the perturbations' excited state after the sharp feature. Due to the difficulty of this calculation and the unwieldiness of the Bogoliubov coefficients, we do not present them in this text but have uploaded them as supplementary material in both Julia and Mathematica code so that further study is easily available to the community. They are available at \url{https://github.com/rjrosati/3field-sharp-feature}.

\item Compared to the two-field case, the three-field excited state has a very different structure for the modes at scales near the feature scale and in its superhorizon. As seen in Figure \ref{fig:bog_comparison}, the two-field well-resolved peak in $k$ vanishes as the torsion increases, giving way to an approximately 1/$k$ power law growth towards the superhorizon.
Similarly, the approximately equal amplitudes of the Bogoliubov coefficients split into a strong hierarchy at high torsion, with the mass parameters choosing the dominant coefficients. Said differently, as the number of fields ${\cal N}$ grows, the contributions from the fields are not the same, and $\Omega_{\rm{GW}}$ does not grow as ${\cal N}^4$.

\item This excited state structure significantly enhances the contribution to the stochastic gravitational wave background (SGWB) during the inflationary era compared to the radiation era. Interestingly, despite the differences in the Bogoliubov coefficients between two-field and three-field scenarios, the frequency of oscillations and the envelope of the signals approximately match the two-field signal, unless $\rho = \Omega/\tau \ll 1$. Several such spectra can be observed in Figure \ref{fig:inflationary-sgwb}. 

Thus, the findings presented in this work support the claim made in \cite{Fumagalli:2021mpc} that ``the principal properties of $\Omega_{\rm{GW}}^{\rm{inf}}$ are independent of the precise shape of $|\alpha(k)|^2$.'' This work expands the class of Bogoliubov coefficients that lead to this universal behavior. The uniqueness of the $\Omega_{\rm{GW}}^{\rm{inf}}$ profile, especially near its peak, makes it a robust tool from the phenomenological perspective.

\end{itemize}

The significant transient growth and decay in the isocurvature modes would typically be unobservable after the end of inflation; however, in our case, they are reflected in the enhancement of the inflationary-era signal. When the feature is not excessively strong, we create a new window of parameter space for SGWB detectability, as only minor adiabatic enhancements are needed to generate a detectable inflationary-era SGWB.

However, when the turn is strong and the maximum $Q_s/\Mpl \sim 1$, this enhancement can be large enough to doubt perturbation theory.  Lattice simulations have found that backreaction plays an important role in geometrical destabilization and several other situations. However, to our knowledge, no such simulations have studied sharp features that correspond to brief violations of perturbation theory.
We suspect that these corrections will further affect the shape of the inflationary-era SGWB and likely reduce the effective amplitude of the isocurvature perturbations when linear theory predicts them to be large.
We stress, however, that due to the integrated nature of the signal (cf. \eqref{eq:tensor_power_spectrum}), even brief transient growths during inflation can be visible in the inflationary-era SGWB.

The background dynamics in \eqref{eq:turn_profile} are unrealistic, and sharp features from concrete models (e.g. those in \cite{Braglia:2020taf, Bhattacharya:2022fze}) often contain several subsequent peaks in the turn rate, affecting the structure of the feature and some of the details of its phenomenology. In Appendix \ref{sec:bch}, we study one particular concrete realization of a model with a sharp feature, and find that it has qualitatively similar results to the toy model studied in this work. The SGWBs from broad features and constant turns in three-field models have been studied in \cite{Christodoulidis:2023eiw}.

We hope that future work will solve these issues so that any future SGWB search can have accurate predictions for a wide array of inflationary features.

\section{Acknowledgments}
We would like to thank A. Achúcarro, P. Christodoulidis, J. Fumagalli, E. McDonough, G. Palma, L. Pinol, S. Renaux-Petel, K. Turzynsky, and I. Zavala for their valuable comments and discussions after presenting this work at the following workshops: Simons Center’s ``Multifield Cosmology: Inflation, Dark Energy and More”, MIAPbP’s ``Quantum Aspects of Inflationary Cosmology”, and ESI’s ``The Landscape vs the Swampland.”
 RR is supported by an appointment to the NASA Postdoctoral Program at the NASA Marshall Space Flight Center, administered by Oak Ridge Associated Universities under contract with NASA. The work of VA and SP was supported in part by the National Science Foundation under Grant No. PHY– 2210562.

\bibliography{refs}{}

\bibliographystyle{JHEP}

\appendix

\section{An example concrete 3-field model with a sharp feature}
\label{sec:bch}
In this section we write down an explicit 3-field model containing a sharp feature, and qualitatively compare it to our results presented in this paper based on top hat turn profiles \eqref{eq:turn_profile} and a very specific mass parametrization \eqref{eq:nonkinetic_masses}.

We consider a model generalized from ``Model 1'' presented in \cite{Braglia:2020taf}, a sort of waterfall mechanism.
Unlike the rest of this work, we do not express the fields in the covariant kinetic basis but directly in the nonlinear sigma model Lagrangian basis, as in \eqref{eq:action}.

We name the fields $\vec\phi \equiv \{\sigma, \chi, \theta \}$ and express the potential and field space metric as
\begin{align}
    V(\sigma,\chi,\theta) &= \Lambda \left[ C_\sigma \left(1-\exp\left(- \frac{\sigma^2}{\sigma_f^2} \right) 
    \right) + (1-C_\chi \chi^2) + (1-C_\theta \theta^2) \right] \\
    G_{ij}(\sigma,\chi,\theta) &= \begin{pmatrix} 1 & 0 & 0 \\
    0 & (1+\xi_{\sigma_f} \sigma/\sigma_f)^2 & 0 \\
    0 & 0 & (1+\xi_{\sigma_f} \sigma/\sigma_f)^2
    \end{pmatrix}
    \label{eq:bch3field}
\end{align}
This is a natural generalization of the two-field Model 1, obtained by simply ``copying'' the form of the $\theta$-field potential and metric contribution into the $\chi$ equivalents. Recovering the original two-field model is as simple as setting $C_\chi = \chi = 0$ and using the $(\sigma,\theta)$ $2\times2$ block of $G_{ij}$.

We take as initial conditions a generalization of the parameters ``A'' of \cite{Braglia:2020taf}: $\vec{\phi}_0 = \{2.12,8\times10^{-3},9.84\times 10^{-3}\}$ and zero initial velocities, with the model parameters $\{ \xi_{\sigma_f}, \sigma_f, C_\sigma, C_\theta, C_\chi \} = \{ 3.327, 0.775, 10.0, 0.0376, 0.01 \}$. We solve for the fields' background evolution, perturbations, and radiation-era induced gravitational waves using \texttt{Inflation.jl} \cite{Inflationjl}.

\begin{figure}[t]
    \centering
    \includegraphics[width=0.9\textwidth]{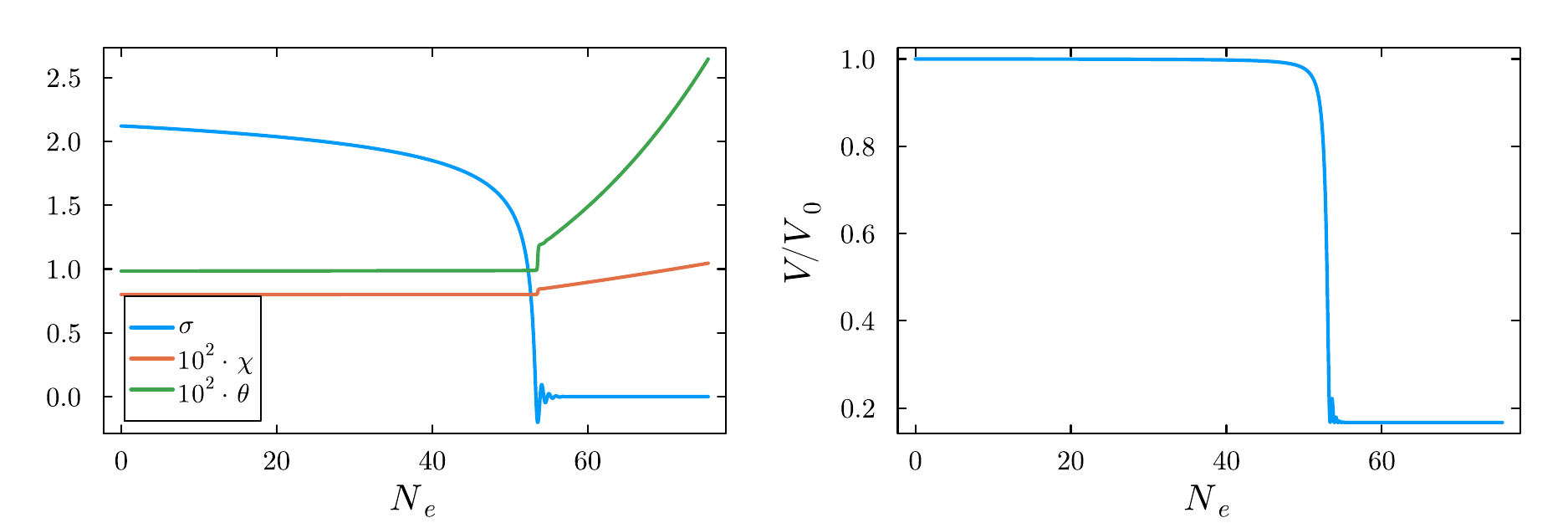}
    \caption{We show the background trajectory in the model \eqref{eq:bch3field}, as well as the value of the potential along the trajectory. The fields experience a ``waterfall'' approximately 21 e-folds before the end of inflation, where the evolution transitions from the $\sigma$ direction to the other fields. In this case, we manually terminate inflation to put the feature at the correct scales for LISA.}
    \label{fig:bch3fieldtraj}
\end{figure}

The evolution of the fields and the potential along the inflationary trajectory is visible in Figure \ref{fig:bch3fieldtraj}. The fields undergo a waterfall-type feature, where initial motion along $\sigma$ decays in violent oscillations along the uphill direction of a saddle point. After relaxing into one of the downhill directions, the fields continue in a new slow-roll slow-turn phase along a mixture of $\chi$ and $\theta$.

\begin{figure}[h]
    \centering
    \includegraphics[width=0.9\textwidth]{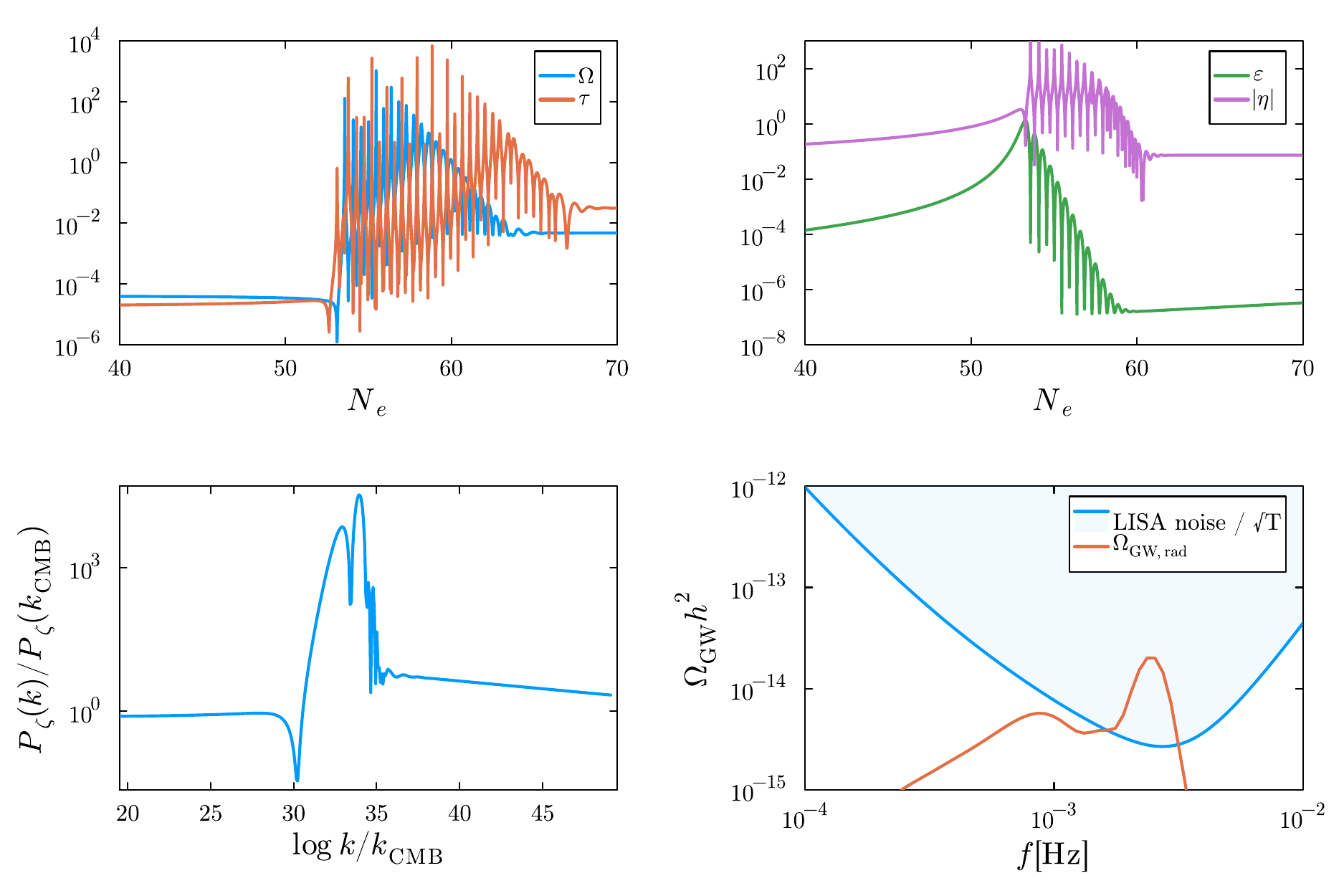}
    \caption{We show the (starting top left) turning rates, slow-roll parameters, scalar power spectrum and radiation-era scalar-induced gravitational waves for the model \eqref{eq:bch3field}. Note that we have synchronized the x-axes of the left column of plots, so that the $k$-modes of $P_\zeta$ are aligned with the e-fold number at which they exit the horizon.
    This waterfall-type feature generates periodic spikes in both $\Omega$ and $\tau$ for a period of $\sim 10$ e-folds before stabilizing into a new slow-roll slow-turn phase of inflation.
    Nonetheless, the enhancement to $P_\zeta$ is localized around modes exiting the horizon within a few e-folds of the first turn rate spikes. This power spectrum contains some oscillations similar to the top hat profile we've studied in this work, although significantly smoother, presumably due to the superposition effects of several sourcing events.
    This power spectrum then generates a theoretically detectably-loud $\Omega_\mathrm{GW, rad}$ at LISA frequencies. The spectral profile of the gravitational waves remains qualitatively similar to the top hat case, but lacks the fine oscillations around the peak.
    Unfortunately, \texttt{Inflation.jl} is a transport-method solver and cannot compute the inflationary-era gravitational waves that should be produced by this feature using the methods developed in this work.
    }
    \label{fig:bch3fieldfeature}
\end{figure}
In Figure \ref{fig:bch3fieldfeature}, we can see the very large spikes induced in the turn rates as the fields oscillate in the saddle. The strongest spikes exceed instantaneous values of $\Omega,\tau \gtrsim 10^3$ and last small fractions of an e-fold.
These turning spikes continue for $\sim 10$ e-folds after they begin and the fields finally relax.
This feature is not slow-roll, as $\epsilon\sim 1$ and $|\eta|\sim 10^2$ during the most violent era.

Despite these dynamics barely resembling the toy model of this work (c.f. Figure \ref{fig:turn_profile}), the induced power spectrum remains remarkably qualitatively similar. The first modes to show an enhancement are a few e-folds before their horizon exits when the waterfall occurs, and the last modes to show an enhancement only exit $\lesssim 5$ e-folds later.
The feature only strongly sources modes at its beginning, and the strong oscillations in the saddle do not contribute to a long period of enhancement in $P_\zeta$.
The oscillations in $P_\zeta(k)$ are much more suppressed than in the top hat model, showing only a few oscillations and with much lower amplitude than $\mathcal{O}(1)$.
We presume this is due to multiple distinct sourcing events contributing to $P_\zeta$ at different times and smoothing out the oscillations we would expect from a single feature (c.f. Figure \ref{fig:turn_profile_Pz}).

The radiation-era gravitational waves produced by this model are in the LISA band but unlikely to be directly observable as written (SNR $\sim 1$).
Nevertheless, it is interesting to focus on their qualitative similarity to the top hat case. The ``shouldered-peak'' structure remains visible and prominent, although the oscillations visible in the top hat case are absent, due to their reduced amplitude in the power spectrum.

Unfortunately, we are unable to (easily) compute the inflationary-era induced gravitational waves numerically for this model.
\texttt{Inflation.jl} is a transport-method-based inflationary solver \cite{Dias:2015rca} and can only directly compute two-point correlation functions of the inflationary perturbations.
This means that Eq. \eqref{eq:tensor_power_spectrum}, which relies on direct access to the $Q_i$, cannot be expressed in terms of the internal variables accessible to transport method solvers.
A future numerical study of this model, another model generalized from \cite{Braglia:2020taf}, or any other multi-field inflationary model would certainly be interesting to perform in future work.

We maintain our claim that the structure of the inflationary-era scalar-induced gravitational waves should remain qualitatively similar to the top hat case studied in this work. The inflationary-era case is much less sensitive than the radiation-era to details of the feature, requiring only an excited state of the perturbations to feed its sourcing, with the shape largely then dictated by the structure of the Green's functions, kernel, and causality constraints.

\end{document}